\documentclass[prl,aps,twocolumn,superscriptaddress,amsmath,amssymb,showpacs]{revtex4}

\usepackage{epsfig}
\usepackage{graphicx}
\usepackage{dcolumn}
\usepackage{bm}

\begin{document}

\title{      Stripe Phases in Layered Nickelates}

\author{     Krzysztof Ro\'sciszewski}
\affiliation{Marian Smoluchowski Institute of Physics, Jagellonian
             University, \\ Reymonta 4, PL-30059 Krak\'ow, Poland }

\author{     Andrzej M. Ole\'s }
\affiliation{Marian Smoluchowski Institute of Physics, Jagellonian
             University, \\ Reymonta 4, PL-30059 Krak\'ow, Poland }
\affiliation{Max-Planck-Institut FKF, Heisenbergstrasse 1,
             D-70569 Stuttgart, Germany }

\begin{abstract}
To describe quasi two-dimensional nickelates we introduce an
effective Hamiltonian for $e_g$ electrons which includes the
kinetic energy, on-site Coulomb interactions, spin-spin and
Jahn-Teller (static) terms. The experimental stripe phases are
correctly reproduced by the model. The mechanisms responsible for
stripe formation are different than those reported in cuprates and
manganites.
\\
{\it Published in: Acta Phys. Polon. A \textbf{121}, 1048 (2012).}
\end{abstract}


\maketitle

Strong electron correlations are responsible for numerous
interesting properties of doped transition metal oxides. In these
systems coexisting charge, magnetic and orbital order emerge from
competition of different type of electronic energy and the
coupling with the lattice \cite{Ole10}. Among them the phenomenon
of stripes is common --- it plays an important role in cuprates
\cite{Voj09} and occurs also in nickelates \cite{Tra94} and
manganites \cite{Dag01}. The stripes in cuprates arise from the
competition between kinetic energy of doped holes and magnetic
energy of ordered spins \cite{Voj09}. The evolution of metallic
stripes under increasing hole doping and their spectral properties
could be described within a purely electronic model \cite{Fle01}.

In contrast, the origin of stripe phases which involve orbital
order is more subtle. The kinetic energy is here
anisotropic and the orbital states easily couple to the lattice.
The extension by Jahn-Teller (JT) interactions is necessary in the
models describing doped manganites \cite{rosc} and nickelates
\cite{Hot04}. The stripes in monolayer nickelates
La$_{2-x}$Sr$_{1+2x}$NiO$_4$ were observed in a range of doping
$0.289<x\le 0.5$ \cite{Tra94}.
They occur as charge walls in an antiferromagnetic (AF) phase and
were described within the electronic model in a broad range of
doping $x<0.4$ \cite{Rac06}, while 
the electronic structure calculations reproduced them at
$x=1/3$ and 1/2 \cite{Yam07,Fre08}. Here we
describe stripe phases within
an effective model featuring only $e_g$ electrons at Ni sites
renormalized by oxygen ions. At each
site the local basis is given by two $e_g$ orbitals
symmetry, i.e., $x\equiv x^2-y^2$ and $z\equiv 3z^2-r^2$ orbitals.

The degenerate Hubbard Hamiltonian ${\cal H}$ for two $e_g$
orbital states contains several terms,
%
\begin{equation}
{\cal H} = H_{\rm kin}+H_{\rm cr}+H_{\rm int}+H_{\rm JT} .
\end{equation}
The kinetic part $H_{\rm kin}$ is given by $(dd\sigma)$ element $t$,
\begin{eqnarray}
H_{kin}&=& -\frac{1}{4} t \sum_{ \{i j\} ||ab, \sigma} \Big\{
 (  3 d^{\dagger}_{i x \sigma}  d_{j x \sigma}  +
     d^{\dagger}_{i z \sigma}  d_{j z \sigma} ) \nonumber \\
& & \hskip 1.7cm \pm \sqrt{3} ( d^{\dagger}_{i x \sigma}  d_{j z \sigma}
+ d^{\dagger}_{i z\sigma}  d_{j x \sigma} ) \Big\},
\end{eqnarray}
and contains the hopping between $x$ and $z$ orbitals at neighboring
sites $\{ij\}$ which changes sign between bonds along $a$ and $b$ axis.
The creation operators $d^{\dagger}_{i \mu\sigma}$ correspond to
electron in orbital $\mu = x,z $, with spin $\sigma=\uparrow,\downarrow$
located at the site $i$. The kinetic energy is supplemented by orbital
splitting $E_z>0$ due to crystal field
\begin{equation}
H_{\rm cr}= -\frac12 E_z\sum_{i \sigma}(
n_{ix\sigma}-n_{iz\sigma}) =- E_z\sum_i\tau_i^z.
\end{equation}
$H_{\rm int}$ stands for on-site Coulomb interactions between $e_g$ 
electrons and was used before for monolayer, bilayer and 
cubic manganites \cite{rosc} and does not need to be reproduced here; 
it is parametrized by intraorbital 
Coulomb element $U$ and Hund's exchange $J_H$.

The simplified JT interaction $H_{\rm JT}$ is:
\begin{eqnarray}
\label{HJT}
H_{\rm JT} &=& 2g_{\rm JT} \sum_i\left\{\left( \frac12 Q_{1 i}(2 -
n_{i}) + Q_{2 i} \tau_i^x+ Q_{3 i} \tau_i^z  \right)\right\} \nonumber \\
&+& \frac{1}{2}\,K \sum_i \Big\{2 Q_{1 i}^2 +Q_{2 i}^2+Q_{3i}^2 \Big\},
\end{eqnarray}
where $n_i=\sum_{\alpha\sigma}n_{i\alpha\sigma}$ is $e_g$ electron 
density operator and the pseudospin $\tau=1/2$ operators are defined as follows:
\begin{equation}
\tau_i^x\!=\!\frac12\sum_\sigma
( d^{\dagger}_{i x \sigma}  d_{i z\sigma} +
d^{\dagger}_{i z \sigma}  d_{i x\sigma} ), \hskip .3cm
\tau_i^z\!=\!\frac12(n_{ix}-n_{iz}).
\end{equation}
The JT term  includes three modes,  $\{Q_{1 i}$,  $Q_{2 i}, Q_{3
i}\}$ which denote standard $e_g$ static deformations  of the
octahedron around site $i$. Note that for dynamic JT effect each two
neighboring Ni$^{2+}$ ions share one oxygen and thus the
JT distortions around them are not independent.

We performed  Hartree-Fock (HF) computations on a finite $6\times 6$ 
cluster (with periodic boundary conditions) filled by $N$ electrons and 
determined the ground state by comparing energies of several converged 
HF states. Following them the electron
correlations were studied using Local Ansatz (LA) method which
implements the leading on-site electron correlations \cite{LA}. Due to
high density of $e_g$ electrons per site $n=2-x$ the correlation 
energies were found to be much larger here than those reported  
before for cuprates \cite{Gor99} or manganites \cite{rosc}, 
but they do not modify the stability of the HF ground state.

\begin{figure}[t!]
\centerline{
\includegraphics[width=5.7cm]{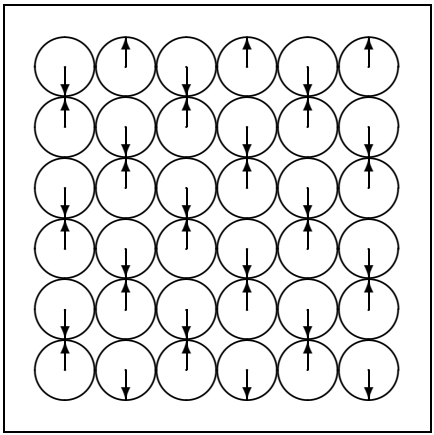}
}
\caption{The antiferromagnetic ground
state obtained for La$_2$NiO$_4$ (at $x=0$). Each circle represents
$e_g$ electron density with its diameter corresponding to one-half
of density; the arrow length --- one-half of the $e_g$ spin; the
horizontal bar length (here absent) --- charge density difference
between $x$ and $z$ orbitals (longest bar to the right/left
corresponds to pure $x/z$ orbital state). 
Parameters (in eV): $t=0.6$, $J_H=0.9$, $E_z=-0.6$; 
$U=8 t_0$, $g_{\rm JT}$ = 3.0 eV/\AA, $K=$ eV/\AA$^2$.}
\label{fig:00a}
\end{figure}

In the undoped La$_2$NiO$_4$ ($N=2\times 6^2$ corresponds to $n=2$
electrons per site) we found a uniform charge distribution in the
ground state, with equal orbital electron densities $n_x=n_z=1$,
corresponding to a high-spin $S=1$ state at each Ni$^{2+}$ ion.
The ionic spins $S=1$ are coupled by the AF interaction which
follows from the Hamiltonian ${\cal H}$ Eq. (1) in a similar way as 
it does for $s=1/2$ spins in cuprates. This explains the origin of 
$G$-type AF order ($G$-AF) shown in Fig. 1.

\begin{figure}[t!]
\centerline{
\includegraphics[width=5.7cm]{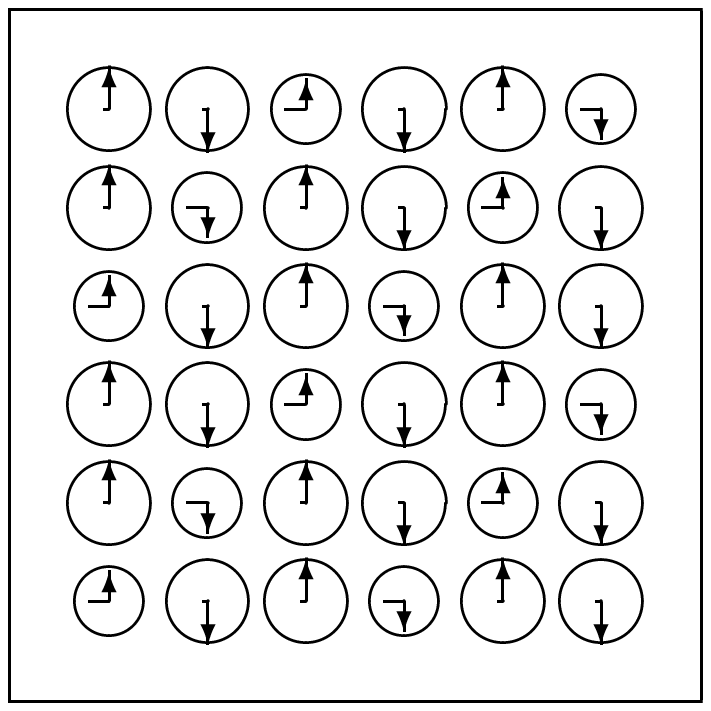}
}
\centerline{
\includegraphics[width=5.7cm]{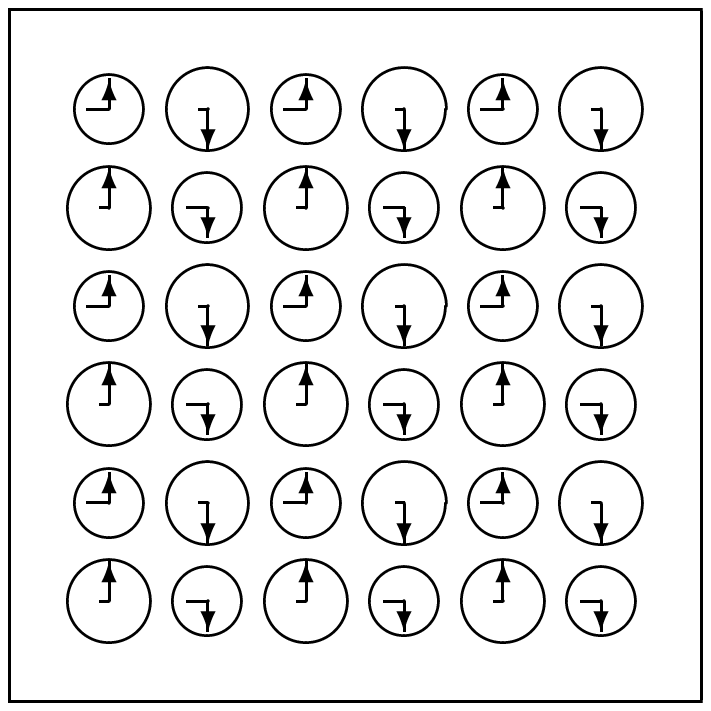}
}
\caption{Two stripe phases with $C$-AF spin order (FM vertical and AF
horizontal order) obtained for doped nickelates at doping $x=1/3$ 
(top) and $x=1/3$ (bottom).
Upper panel --- charge minority sites form diagonal 
(11) stripe boundaries (each third line).
Lower panel --- Perfect checkerboard crystal-like order given by
charge majority/minority sublattices. In both cases holes are doped 
predominantly into $x$ orbitals. Legend and parameters as in Fig. 1. }
\label{fig:12a18a}
\end{figure}

Next we investigate the ground states for decreasing electron number 
per site $n=2-x$. At low doping $0.1<x<0.3$ we obtained within the 
present approach isolated polaronic states (not shown). Diagonal 
stripe phases are found for doping $x=1/3$ and $x=1/2$, 
in agreement with experimental observations \cite{Tra94} 
--- they are shown in Figs. 2 and 3. They are not like the stripes in 
cuprates or manganites. In cuprates charge order coexists with the 
modulation of AF order between alternating AF domains, separated by 
nonmagnetic and half-filled domain walls. The JT 
coupling does not contribute here as the hole sites are inactive.
In contrast to cuprates, Ni ions always carry magnetic 
moments and contribute to the magnetic order within a single domain
of the $C$-AF phase. Note also that the present stripe phases are
insulating, while the ones in cuprates are metallic \cite{Fle01}.
   
In manganites the situation is more complex due to possible
presence of orbital order and due to extra Hund's exchange
coupling of $e_g$ electrons with $t_{2g}$ core spins $S=3/2$
\cite{Dag01}. Here several competing mechanisms enter on
equal footing and distinct stripe patterns with the true long
range order are not a rule. As a result, the magnetic order of the
ground state was found to be very sensitive to the precise values
of the model parameters \cite{Dag06}. However, at $x=1/2$ doping one 
finds a robust instability for monolayer manganites toward two 
sublattices with charge majority and minority sites, see also [8a]. 
The JT effect stabilizes here the CE magnetic order accompanied by
checkerboard charge order, with the JT distortions contributing only 
on charge majority sites.
%

\begin{figure}[t!]
\centerline{
\includegraphics[width=5.7cm]{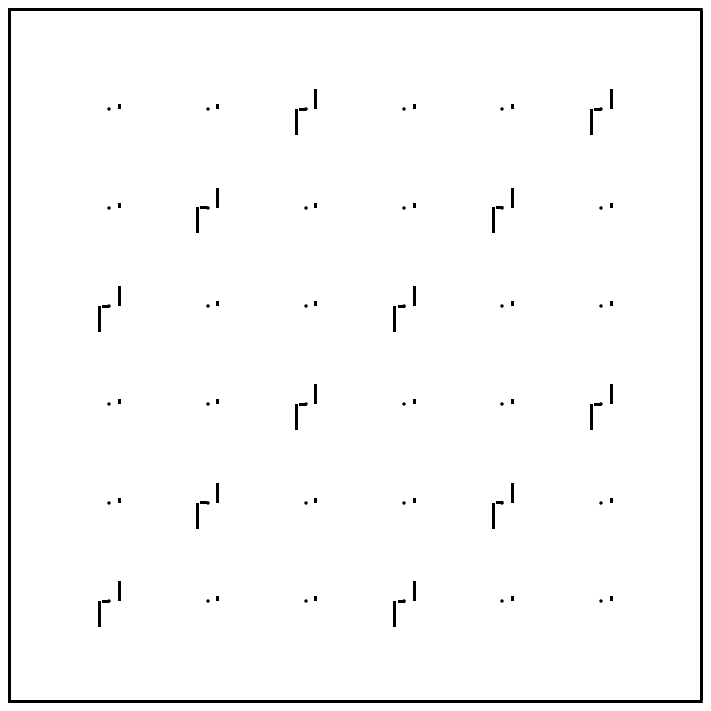}
}
\centerline{
\includegraphics[width=5.7cm]{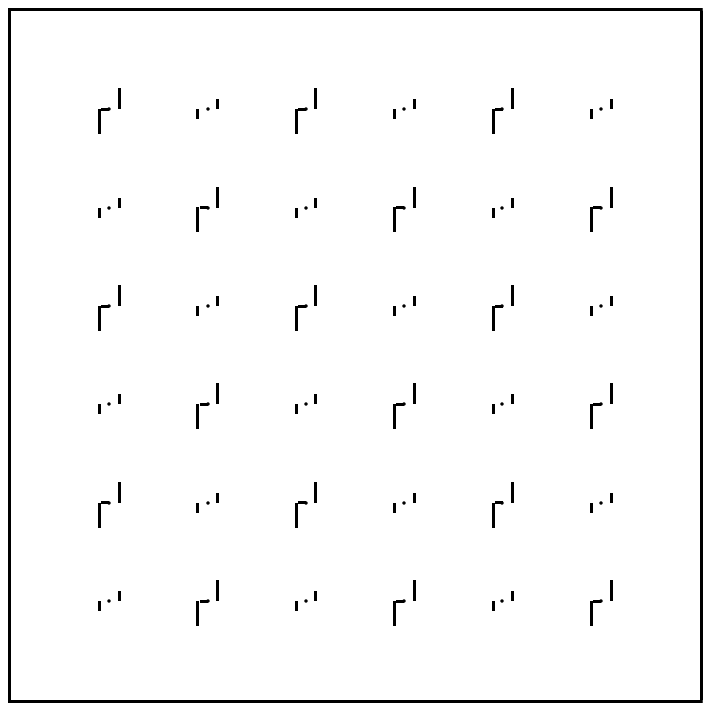}
} \caption{ Jahn-Teller distortions corresponding to the stripe
phases shown in Fig. 2: top --- $x=1/3$, and bottom $x=1/2$ 
doping. The distortions $Q_{2i}$ (in \AA, scaled by a factor of
2) are shown as vertical bars drawn slightly to the left of each
site and $Q_{3i}$, by bars to the right of each site (isotropic
mode $Q_{1i}$ is absent). Legend and the values of parameters as
in Fig. 1. } \label{fig:12b18b}
\end{figure}

In monolayer nickelates one finds the opposite situation: The main
difference in the ionic picture is that in the undoped substance 
there are two $e_g$ electrons per site, and hole doping generates
JT active sites, where a single electron couples to JT distortions 
$Q_{2i}$ and $Q_{3i}$, see Fig. 3. This coupling is crucial for 
stabilizing ordered stripe structures at $x=1/3$ and $x=1/2$ doping, 
shown in Fig. 2. In effect one can say that the kinetic energy of 
doped holes is suppressed and the competition between magnetic and 
kinetic energy along the charge minority walls is quenched, 
in agreement with double exchange \cite{Dag01}. 

The above ionic picture used to interrpret both stripe phases is 
not far from the actual charge distribution obtained in the present
computations. We have found $n_1=1.82$, $n_2=1.37$ ($n_1=1.66$, 
$n_2=1.34$) at charge majority/minority sites at $x=1/3$ ($x=1/2$) 
doping. As quantum fluctuations are suppressed in the HF method, this 
corresponds to magnetic moments $m_1=\pm 0.89$, $m_2=\pm 0.67$
($m_1=\pm 0.81$, $m_2=\pm 0.62$) in both phases.

Doping occurs predominantly in $x$ orbitals, where also the magnetic 
moments are suppressed. The density in $x$ orbitals varies between 
$n_x=0.47(0.85)$ and $n_x=0.45(0.73)$ for the stripe phases at $x=1/3$ 
and $x=1/2$ doping, while density in $z$ orbitals is $n_z\ge 0.89$ in 
all cases. For this charge distribution both $Q_{2i}$ and $Q_{3i}$ JT 
modes are active at charge minority sites, see Fig. 3. We suggest that 
JT distortions govern the stripe formation in monolayer 
nickelates, and stabilize the obtained charge alternation accompanied 
by large JT distortions at charge minority sites. We have found that 
the doped sites can order at doping 
$x=1/3$ and appear as diagonal stripes (each third line). 
Similarly, they give diagonal stripes (each second line) for $x=1/2$ 
doping stabilized by the JT distortions. Indeed, when the JT coupling 
is turned off (at $g_{\rm JT}=0$), the stripes do not form. Strong
electron correlations are also important and we have verified that 
similar stripe phases to those shown in Figs. 2 and 3 
occur for a stronger on-site Coulomb repulsion $U=12t$ in the present 
model Eq. (1), only the charge modulation is somewhat enhanced.

In summary, we identified the coupling of $e_g$ electrons at sites
doped by holes to local Jahn-Teller distortions as the microscopic
origin of diagonal stripe phases in monolayer nickelates. The 
presented results elucidate fundamental difference between stripe
phases in doped nickelates, cuprates and manganites, as the ions 
active for the Jahn-Teller effect play a different role in each case.

\section*{Acknowledgments}
We acknowledge support by the Polish Ministry of Science and
Higher Education under Project No. N202~104138. A.M. Ole\'s
acknowledges also support by the Foundation for Polish Science
(FNP).

\end{document}